\documentclass{ws-procs9x6}

\begin{document}


\title{
\vspace{-5\baselineskip}
\begingroup
\footnotesize\normalfont\raggedleft
\lowercase{\sf hep-th/0402023} \\ 
ILL-(TH)-04-01\\
\vspace{\baselineskip}
\endgroup
Mathematical Aspects of D-branes\footnote{
\uppercase{T}o appear in the \uppercase{P}roceedings of the 3rd 
\uppercase{S}ymposium on \uppercase{Q}uantum
\uppercase{T}heory and \uppercase{S}ymmetries (\uppercase{QTS}3),
\uppercase{C}incinnati, \uppercase{O}hio,
10--14 \uppercase{S}ept 2003 ---
\copyright\
\uppercase{W}orld \uppercase{S}cientific.}}

\author{Eric Sharpe}
\address{Physics Department\\University of Illinois\\
Urbana, IL  61801\\E-mail:  ersharpe@uiuc.edu}  

\maketitle

\abstracts{
In this lecture we review recent work on
describing D-branes with nonzero Higgs vevs in terms of sheaves,
which gives a physical on-shell D-brane interpretation to more sheaves
than previously understood as describing D-branes.  
The mathematical ansatz for
this encoding is checked by comparing open string spectra between
D-branes with nonzero Higgs vevs to Ext groups between the
corresponding sheaves.  We illustrate the general
methods with a few examples.
}


\section{Introduction}

Several years ago, Kontsevich \cite{kont} proposed a 
formulation of mirror symmetry in terms of derived categories.
Motivated by Kontsevich's work together with
physical pictures of tachyon condensation \cite{sen} and K-theory
\cite{edkthy},
it was originally proposed in
\cite{dercat} that Kontsevich's proposal could be physically realized
via off-shell states in the open string B model, in which objects in the
derived category (represented by complexes of locally-free sheaves)
are represented by D-brane/anti-D-brane configurations, with maps between
sheaves represented by tachyons, and localization on quasi-isomorphisms
realized by renormalization group flow.  Some evidence was given for this
proposal by considering T-duality arguments.

This proposal lay dormant until the work of \cite{doug,paulalb},
which popularized these ideas, and also introduced the notion
of ``$\pi$-stability.''   

However, to this day, there are still many open problems in 
identifying (off-shell)
open string B model
boundary states with objects in the derived category of coherent
sheaves.

The most basic problem is that of confirming that 
the mathematical operation of localizing on quasi-isomorphisms really is
realized physically via renormalization-group flow.  Since in general
cases it is all but impossible to see the entire RG flow directly,
one must perform indirect tests; however, until the work of
\cite{ks,kps,cks} (see \cite{melec,katzhere} for a review) 
no technology existed for confirming any of the predictions of the
derived category program directly in BCFT.  
Interesting subtleties were also uncovered in \cite{ks,kps,cks};
for example, the Freed-Witten anomaly \cite{freeded} complicates the dictionary
between D-branes and sheaves, and plays a vital role in deriving
Ext groups as open string states. 

A second issue in the derived categories program  
has been known for almost eight years, even predating
\cite{dercat}.  This issue is the face that the GSO projection
tells us that degrees of Ext group elements cannot match
$U(1)_R$ charges in general, except in the special case of open strings
connected D-branes to themselves, contradicting naive assumptions.  
This mismatch is manifest in BCFT but very obscure in massive nonconformal
models, and as such mismatches can have drastic implications for RG flow
arguments, not only is understanding it important, but its existence
underlines the importance of checks of RG flow arguments, of the form just
mentioned.

Third, the open string analogue of the B model Calabi-Yau condition
places a constraint on the possible morphisms of a derived category
that can be realized physically by open strings.  Although it appears
that all objects of a derived category can be represented by
off-shell open string states, it seems that not all the morphisms in
a derived category can be represented physically.   

Finally, 
we now believe we understand that sheaves of the form $i_* {\mathcal E}$,
where $i$ is an inclusion, model on-shell D-branes, but do other coherent
sheaves model on-shell D-branes, and if so, which on-shell D-branes?
This question was partially answered in \cite{tomasme,dks}, and this short 
note is devoted to reviewing 
the answer to this particular question.

Very briefly, some sheaves not of the form $i_* {\mathcal E}$ have
a physical interpretation as D-branes with Higgs vevs.
`Easy' Higgs vevs merely move the support of the sheaf,
but {\it e.g.} nilpotent Higgs vevs deform the sheaf nontrivially to
sheaves not of the form $i_* {\mathcal E}$.
In a non-topological theory, nilpotent Higgs vevs are often excluded
by D-terms (with the prominent exception of orbifolds,
where the nilpotents map out exceptional divisors, as seen in
\cite{dgm}), but in a topological theory, there is no such constraint.

\section{Description of D-branes with nonzero Higgs vevs in terms of sheaves}

A sheaf of the form $i_* {\mathcal E}$, where $i: S \hookrightarrow X$,
describes a D-brane on $S$ with gauge bundle ${\mathcal E} \otimes
\sqrt{K_S^{\vee}}$, and with vanishing Higgs vevs, which clearly is not
the most general case.

To describe D-branes with nonzero Higgs vevs in terms of sheaves we can use 
a mathematical prescription 
that has appeared previously in \cite{del}.  In a nutshell, one uses the
Higgs vevs to define a deformation of the action of the ring of
algebraic functions on the sheaf.  Coordinates describing directions
normal to the support, that would ordinarily annihilate the module 
corresponding to the sheaf, instead are taken to act nontrivially
on the module, with action defined by the Higgs fields.  The result
is a new sheaf, living in the total space of the normal bundle to the
support of the old sheaf.  We shall
see this explicitly in examples below.

A mathematical ansatz for encoding Higgs vevs in sheaves is not
enough to guarantee that the resulting sheaves have physical content.
To verify that this mathematical ansatz is indeed relevant,
we compare Ext groups between the sheaves generated by this ansatz
to the physical open string spectra, which are calculated using
essentially the methods of \cite{ks}, except that 
nonzero Higgs vevs deform the worldsheet BRST operator \cite{paulalb,dks}
to the form
\begin{displaymath}
Q \: = \: \overline{\partial} \: + \: \Phi^L_i \theta_i \: - \:
\Phi^R_i \theta_i.
\end{displaymath}

In \cite{dks} we prove the
\begin{theorem}
Ext groups between sheaves generated by the mathematical ansatz above always
match open string spectra computed by the BRST operator above.
\end{theorem}

\section{Example:  Single D0 brane on ${\bf C}$}

The simplest nontrivial example consists of a single D0 brane
on the complex line ${\bf C}$.  Such a D0-brane has a single
Higgs field, call it $\Phi$.

The sheaf corresponding to that Higgs vev is straightforward
to work out.  We begin with a skyscraper sheaf supported at the
origin, described by the module ${\bf C}[x]/(x)$, which we
can describe in terms of a single generator $\alpha$ obeying the
relation $x \cdot \alpha = 0$.
Giving the Higgs field a vev deforms the ring action to become
$x \cdot \alpha = \Phi \alpha$ or more simply,
$(x-\Phi)\cdot \alpha = 0$.  In other words, the new module
is ${\bf C}[x]/(x-\Phi)$, describing a skyscraper sheaf shifted
away from the origin, exactly as one would expect.

As a further check, let us compare open string spectra between
the D0-brane above and another D0-brane with no Higgs vevs, to Ext groups
between the corresponding sheaves.
Since the sheaves corresponding to the two D0-branes are skyscraper sheaves
supported at different points, all Ext groups between the sheaves vanish,
so there should be no physical open string states.

Let us check that BRST cohomology gives the same result.
Degree zero states are defined by constants $b$.
Demanding that such states be annihilated by the BRST operator,
which in the present case simplifies to be merely $Q = \Phi \theta$,
tells us that $b \Phi = 0$, so $b=0$ (unless $\Phi=0$), so there are no
BRST-invariant degree zero states.

Degree one states are of the form $b \theta$, for constant $b$.
All these states are annihilated by the BRST operator above;
they are all also in the image of the BRST operator, hence there
are no degree one states.

Thus, just as predicted, there are indeed no physical open string states
between a pair of D0-branes when one has a nonzero Higgs vev.

\section{Example:  D0 branes on ${\bf C}^2$}

For our next example, we shall consider a pair of
D0 branes at the origin of ${\bf C}^2$.
There are two Higgs fields; we shall give them vevs
\begin{displaymath}
\Phi_x \: = \: \left[ \begin{array}{cc}
                      0 & 1 \\ 0 & 0 \end{array} \right], \: \: \:
\Phi_y \: = \: \left[ \begin{array}{cc}
                      0 & 0 \\ 0 & 0  \end{array} \right].
\end{displaymath}
(Note that these Higgs vevs commute, as needed to satisfy F-term
conditions.)

We can find the corresponding sheaf as follows.
Start with a pair of skyscraper sheaves ${\mathcal O}_0^2$ supported
at the origin, describing the original pair of D0 branes
before the Higgs vevs.
Describe that pair of skyscraper sheaves as a module over ${\bf C}[x,y]$
with generators $\alpha$, $\beta$, annihilated by $x$ and $y$:
\begin{displaymath}
x \cdot \alpha \: = \: x \cdot \beta \: = \:
y \cdot \alpha \: = \: y \cdot \beta \: = \: 0.
\end{displaymath}
Create a new module by deforming the ring action as follows:
\begin{eqnarray*}
x \cdot \left[ \begin{array}{c} 
               \alpha \\ \beta \end{array} \right] & = &
\Phi_x \left[ \begin{array}{c} 
               \alpha \\ \beta \end{array} \right] \: = \:
\left[ \begin{array}{c}
       \beta \\ 0 \end{array} \right], \\
y \cdot \left[ \begin{array}{c} 
               \alpha \\ \beta \end{array} \right] & = &
\Phi_y \left[ \begin{array}{c} 
               \alpha \\ \beta \end{array} \right] \: = \:
\left[ \begin{array}{c}
       0 \\ 0 \end{array} \right].
\end{eqnarray*}
This is the module\footnote{Identify $\alpha$ with $1/(x^2,y)$ and
$\beta$ with $x/(x^2,y)$.} ${\bf C}[x,y]/(x^2,y)$, corresponding to a sheaf
we shall denote $D_x$.

We shall check that this sheaf is physically-relevant by comparing
open string spectra between D-branes to Ext groups between sheaves.
For simplicity, let us compute open string spectra between the D0-branes
with Higgs vevs as above, and another pair of D0-branes with vanishing
Higgs vevs.

Mathematically, it can be shown (see \cite{dks} for details) that
\begin{displaymath}
\mbox{Ext}^n_{ {\bf C}^2 } \left( {\mathcal O}_0^2, D_x \right) \: = \:
\left\{ \begin{array}{cl}
        {\bf C}^2 & n=0,2, \\
        {\bf C}^4 & n=1.
        \end{array} \right.
\end{displaymath}
In general, the mathematical computation is easier than the physical
computation -- indeed, this is one of the motivations for 
describing open string states in terms of Ext groups -- but in this 
simple case, it is also straightforward to
do the physical computation, as we shall now outline.

Degree zero states are of the form
\begin{displaymath}
V \: = \: \left[ \begin{array}{cc}
                 a & b \\
                 c & d \end{array} \right].
\end{displaymath}
Demanding that these states be annihilated by the BRST operator
\begin{displaymath}
Q \: = \: \overline{\partial} \: + \: \Phi^L_i \theta_i \: - \:
\Phi^R_i \theta_i
\end{displaymath}
implies the constraint $V \Phi_x = 0$,
which implies that the BRST-invariant states have $a=c=0$,
and so the space of BRST-invariant states is two-dimensional,
matching the Ext group computation. 

Degree one states are of the form
\begin{displaymath}
V \: = \: \left[ \begin{array}{cc}
                 a_x & b_x \\ c_x & d_x \end{array} \right] \theta_x \: + \:
\left[ \begin{array}{cc}
       a_y & b_y \\ c_y & d_y \end{array} \right] \theta_y.
\end{displaymath}
States that are annhilated by the BRST operator have 
$a_y=c_y=0$,
and the image of the BRST operator has the form
\begin{displaymath}
\left[ \begin{array}{cc}
       0 & a \\ 0 & c \end{array} \right] \theta_x
\end{displaymath}
so we see that the space of degree one states has dimension
$6-2=4$, matching the Ext group computation.

Degree two states are of the form
\begin{displaymath}
V \: = \: \left[ \begin{array}{cc}
                 a & b \\
                 c & d \end{array} \right] \theta_x \theta_y.
\end{displaymath}
All of these states are annihilated by the BRST operator,
and states in the image of the BRST operator have $a=c=0$,
so we see that the space of degree two states has dimension $4-2=2$,
matching the Ext group computation.

%

\section{Relevance to D-branes in orbifolds}

As mentioned in the introduction, nilpotent Higgs fields
occur in non-topological theories in orbifolds.
In \cite{dgm} it was shown that classical Higgs moduli spaces
of D-branes in orbifolds encode resolution of quotient spaces.
It is straightforward to check that the exceptional divisors
of those resolutions are described by nilpotent Higgs fields,
which correspond via our construction to $G$-equivariant
nonreduced schemes on the covering space (equivalently,
nonreduced schemes on the quotient stack).
For example, the previous section contained a set of
${\bf Z}_2$-invariant Higgs fields relevant to describing
D0-branes on $[{\bf C}^2/{\bf Z}_2]$, which we saw explicitly
were described by a nonreduced scheme on the covering space ${\bf C}^2$. 

The McKay correspondence, in the form \cite{bkr}, also relates
exceptional divisors to $G$-equivariant nonreduced schemes, via
a different mechanism, not directly involving D-branes.

In effect, the construction we have outlined here closes this
circle of ideas involving the McKay correspondence and the
calculations of \cite{dgm}.  See our paper \cite{dks} for more
details.

\section*{Acknowledgments}

We would like to thank Sheldon~Katz for many useful
conversations and collaborations on this material.


\end{document}